\documentclass[reprint,aps,pre,floatfix,footinbib,superscriptaddress]{revtex4-1}
\usepackage{color}
\usepackage[utf8]{inputenc}
\usepackage[T1]{fontenc}

\usepackage{graphicx}
\usepackage{dcolumn}
\usepackage{bm}
\usepackage{amsmath,amssymb,amsthm} 
\usepackage{makeidx}
\usepackage{amsfonts}
\usepackage{comment}
\usepackage[usenames,dvipsnames]{pstricks}
\usepackage{subfigure}
\usepackage{braket}
\usepackage{float}

\DeclareMathOperator*{\argmax}{argmax}
\newcommand{\Mv}{$\bm{\mathcal{M}}_\tau \,$ }

\begin{document}

\title{Scale-dependent measure of network centrality from diffusion dynamics}

\author{Alexis Arnaudon}
\affiliation{Department of Mathematics, Imperial College London, London SW7 2AZ, UK}
\altaffiliation{\textit{Current address of AA:} Blue Brain Project, École Polytechnique Fédérale de Lausanne (EPFL), CampusBiotech, 1202 Geneva, Switzerland}
\author{Robert L.\ Peach}
\affiliation{Department of Mathematics, Imperial College London, London SW7 2AZ, UK}
\affiliation{Imperial College Business School, Imperial College London, London SW7 2AZ, UK}
\author{Mauricio Barahona}
\email{m.barahona@imperial.ac.uk}
\affiliation{Department of Mathematics, Imperial College London, London SW7 2AZ, UK}

\begin{abstract}
Classic measures of graph centrality capture distinct aspects of node importance, from the local (e.g., degree) to the global (e.g., closeness). 
Here we exploit the connection between diffusion and geometry to introduce a multiscale centrality measure.
A node is defined to be central if it breaks the metricity of the diffusion as a consequence of the effective boundaries and inhomogeneities in the graph. 
Our measure is naturally multiscale, as it is computed relative to graph neighbourhoods within the varying time horizon of the diffusion. 
We find that the centrality of nodes can differ widely at different scales.
In particular, our measure correlates with degree (i.e., hubs) at small scales and with closeness (i.e., bridges) at large scales, and also reveals the existence of multi-centric structures in complex networks.
By examining centrality across scales, our measure thus provides an evaluation of node importance relative to local and global processes on the network. 

\end{abstract}

\maketitle

\section{Introduction}

Identifying central nodes in a network is a topic of wide interest across fields, from finding critical junctions in a road network or important stations in a power grid to establishing which people are best poised to spread (or stop) gossip in a social network~\cite{Borgatti2006,Newman2010a}. 
Since the 1950s, and motivated by heuristics from different application areas~\cite{bavelas1950communication,katz1953new,beauchamp1965improved}, various notions of node importance have been proposed to measure influence in networks. 
These ideas were formalized and extended by several authors as classical graph-theoretical measures of node centrality~\cite{Harary1969}, some of a combinatorial nature (e.g., degree, closeness, or betweenness centrality~\cite{freeman1978centrality}), others with a spectral character (e.g., eigenvector centrality~\cite{bonacich1972}).
Spurred by the current interest in networks, such measures have been reformulated in different contexts, from extended notions of neighbourhood
(e.g., $k$-core~\cite{kitsak2010identification}) to information propagation (e.g., current-flow closeness~\cite{stephenson1989rethinking}). 
Measures based on random walks (e.g., random walk accessibility~\cite{de2014role}, extensions of betweenness~\cite{Newman2005a}, non-backtracking centrality~\cite{martin2014localization}, embeddability in graph cycles~\cite{Schaub2014a}, or subgraph centrality~\cite{estrada2005subgraph}) have also been proposed. 
For a survey of centralities, see the recent monograph~\cite{fouss2016algorithms}.

These different centrality measures capture relevant, yet distinct aspects of node importance. Some centralities are based on a local notion, such as the number of connections of the node (i.e., the degree), whereas others stem from global network properties, such as closeness centrality, which considers the sum of shortest paths with \emph{all} other nodes.
 However, the implicit scale of each centrality measure is rarely obvious. For instance, betweenness centrality measures the participation of a node in the minimal paths between all pairs of nodes in the graph. As such, it might be expected to be as global a measure as closeness, yet its value is computed only from a potentially small fraction of nodes where geodesic paths concentrate. 
Hence the effective scale of betweenness depends on the graph structure and varies for each node. 
The influence of scale on centrality is recognisable in the fact that the classic Katz centrality~\cite{katz1953new} as well as recent measures~\cite{Estrada2010,gurfinkel2019absorbing,alalwan2019melting} contain 
a free parameter that can be tuned to weigh the relative importance of local \textit{vs.}\ global properties.

Here, we introduce a measure of centrality that is intrinsically multiscale by invoking a notion of scale implicit to the graph, i.e., the centrality of each node is computed over the set of nodes that are reachable over a given time horizon $\tau$ by a dynamics taking place on the graph.
Perhaps the simplest of such processes, which can be derived as an approximation to more realistic dynamics, is a diffusive graph dynamics. In that case, the time horizon of the diffusion plays the role of a natural scale factor: as the diffusion probes the surroundings of a node, it becomes sensitive to the presence of \emph{effective boundaries and geometric constraints}, which affect the shape of its growing neighborhood.
Our measure then exploits the links between diffusion and geometry, as in the classic "hearing the shape of a drum"~\cite{kac1966can}, to establish the centrality of a node in terms of its position relative to its scale-dependent neighbourhood.
Such a geometric interpretation allows us to quantify centrality as a measure of how far a node is from effective boundaries and constraints found via diffusive dynamics in the graph.  
This definition recovers classic notions of graph centrality that go from the local to the global, as we illustrate through a series of examples.  

\section{Multiscale centrality}
\subsection{Definition of multiscale centrality: from diffusion to geometry and graphs}

\begin{figure*}[htpb]
    \centering
    \includegraphics[width=0.95\textwidth]{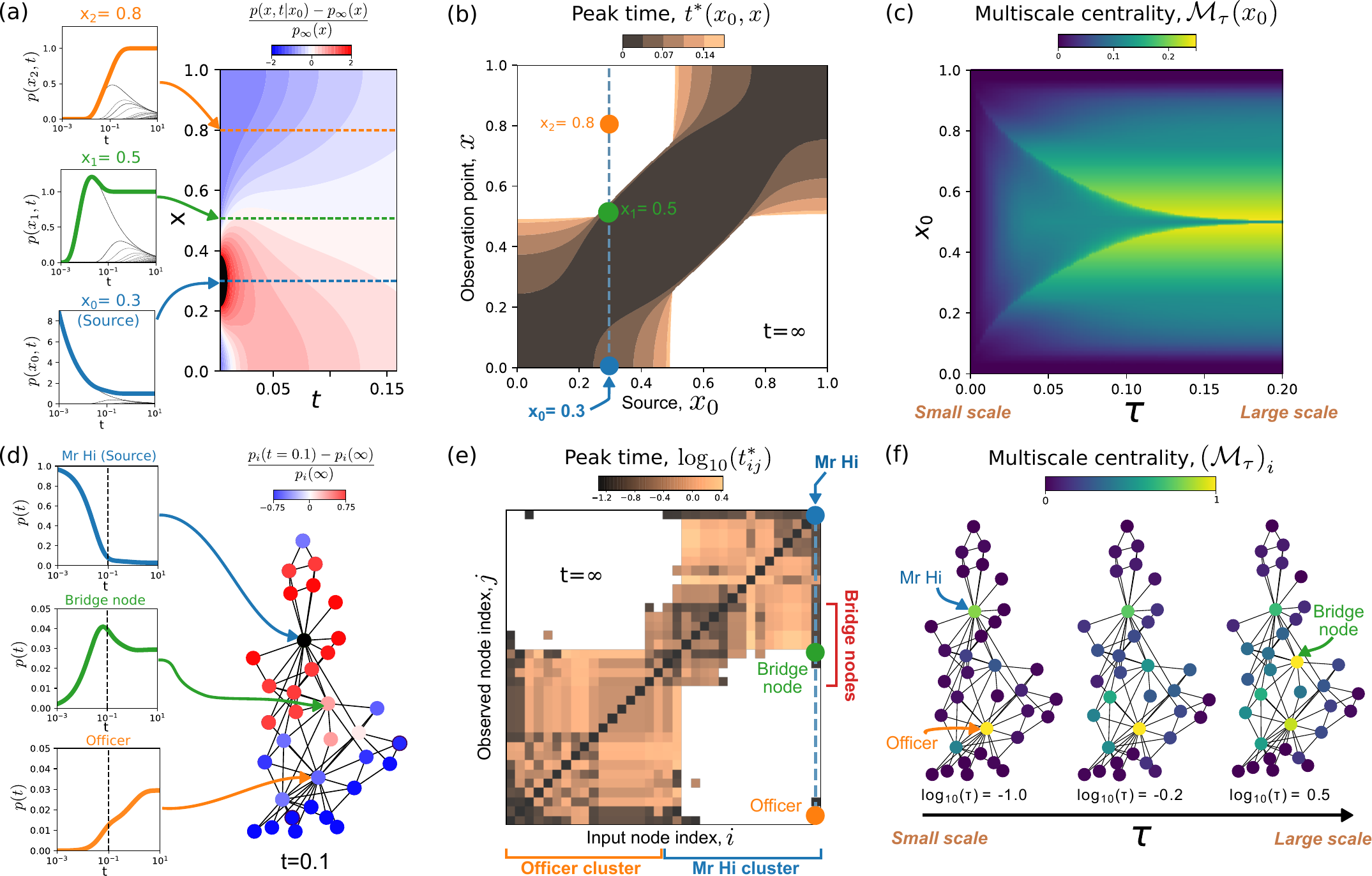}
    \caption{ 
    \textbf{(a)-(c) Diffusion and centrality on the interval $\mathbf{[0,1]}$.} 
    \textbf{(a)} 
    Solution of the diffusion equation~\eqref{diffusion-segment} given a delta function at $x=0.3$. The temporal responses at three points 
    ($x_0=0.3$, $x_1=0.5$, $x_2=0.8$) exhibit presence or absence of peaks. The thin black lines correspond to the Green's functions forming the solution~\eqref{eq:full_solution_line}.
    \textbf{(b)} The peak time function $t^*(x_0,x)$~\eqref{eq:peaktimes} represents the time at which an overshoot is observed at position $x$ for an initial delta function at position $x_0$. The white areas correspond to non-reachable points where $t^*(x_0,x)= \infty$.  
    \textbf{(c)} The multiscale centrality measure $\mathcal{M}_\tau(x_0)$~\eqref{eq:definition_MSC} captures the center of the interval at large scales $\tau$. 
    \textbf{(d)-(f) Multiscale centrality of the Zachary's karate club network.}
    \textbf{(d)} The snapshot of the solution~\eqref{eq:network_solution} of the diffusion starting with an impulse from Mr. Hi after $t=0.1$ highlights the different reachability of different parts of the network.
    As shown in the three insets, only certain nodes experience an overshoot.
    \textbf{(e)} Matrix of peak times $t^*_{ij}$ clustered and ordered according to Ward hierarchical clustering. The peak times correspond well to the standard two-way partition, whilst identifying bridge nodes between the clusters.
    \textbf{(f)} The normalized multiscale centrality~\eqref{eq:norm_MSC} for three values of $\tau$ reveals the importance of bridge nodes at large scales.
}
    \label{fig:main_figure}
\end{figure*}

To motivate our definition, we start with a simple geometric setting and recall how diffusion can be used to compute the center of a compact Euclidean domain relative to its boundaries. Consider the one-dimensional diffusion equation with constant coefficient $D$ with Neumann boundary conditions on a finite domain~$x \in [0,1]$:  
\begin{align}
    \partial_t p(x,t)=  D \, \partial_x^2 p(x,t)\,,
    \label{diffusion-segment}
\end{align}
and let the initial condition be located at $x_0$, $p(x,0|x_0)= \delta(x-x_0)$. The solution can then be computed by the method of images:
{\small \begin{align}
\label{eq:full_solution_line}
p(x,t|x_0) &= \sum_{k=-\infty}^{k = \infty} G_t(x+2k+x_0)+ G_t(x+2k-x_0) \, ,
\end{align}}
where the Green's function is 
\begin{align}
    G_t(x) = (4D\pi t)^{-1/2} \, \exp\left(- x^2/4Dt \right)\, . 
    \label{eq:green}
\end{align}
Figure~\ref{fig:main_figure}(a) shows the solution~\eqref{eq:full_solution_line} for $x_0=0.3$, and the insets highlight the temporal responses at three points: the source point $x_0=0.3$, and two observation points $x_1=0.5, x_2=0.8$.  At the input point $x_0$, the impulse decays monotonically towards the stationary value $p_\infty(x)=1$, whereas for all other points $p(x,t|x_0)$ starts at zero and grows asymptotically towards $p_\infty(x)$. Yet, as Figure~\ref{fig:main_figure}(a) shows, the approach to the stationary value is qualitatively different depending on the distance of the observation point to the source point. Specifically, the temporal response exhibits an \emph{overshooting peak} if the observation point is close enough to $x_0$ (e.g., $x_1$), whereas there is no overshooting if the observation point (e.g., $x_2$) is further away from $x_0$.
To capture this effect, we define the \emph{peak time} at which the maximum overshooting appears at point $x$ when the input was at point $x_0$: 
\begin{align}
\label{eq:peaktimes}
    t^*(x_0,x)= \argmax_{t\in [0, \infty)} p(x, t|x_0) . 
\end{align}
When no overshooting peak is observed at $x$, we adopt the convention $t^*(x_0,x) = \infty$ and denote such points as \emph{non reachable}. 
Figure~\ref{fig:main_figure}(b) shows the peak times~\eqref{eq:peaktimes} for our example, with the non-reachable regions in white.

This dynamical perspective affords us an intrinsic scale defined by the time horizon over which we observe the diffusion, and we define the \emph{reachability within a time horizon $\tau$}: 
\begin{align}
\label{eq:peaktimes_tau}
t^*_\tau(x_0,x)= 
\begin{cases}
\infty & \text{if } t^*(x_0,x)\geq \tau \\
t^*(x_0,x) & \text{otherwise}\, . 
\end{cases}
\end{align}
Clearly, $t^*_\infty(x_0,x)=t^*(x_0,x)$.

It is easy to realize that the reachability of a point follows from both the presence of boundaries in the underlying space and the time horizon over which the diffusive dynamics probes the structure. 
Indeed, if we have an infinite domain $x\in[-\infty,\infty]$, the solution is $p(x,t|x_0) = G_t(x-x_0)$ and the peak time is just $t^*(x_0,x)=(x-x_0)^2/2D, \enskip \forall x_0$, i.e., all points are reachable and the peak time is proportional to the square of the Euclidean distance.
For a finite segment, on the other hand, this simple relationship between scale and distance holds only approximately for points away from the boundaries and over small horizons, and it breaks down as the horizon grows and the dynamics aggregates information about the boundaries.
Due to these boundary effects, the peak time function  $t_\tau^*(x_0,x)$ becomes non-unique and non-convex in its arguments, resulting in violations of the triangular inequality.
Specifically, for a given $x_0$, there can be pairs of points $(x_1, x_2)$ for which we have the following violation:
\begin{equation}
\begin{split}
  \Delta_\tau(x_1,&x_2|x_0) =  \Delta_\tau(x_2,x_1|x_0) \\ 
    & := t_\tau^*(x_0,x_1) + t_\tau^*(x_0,x_2) - t_\tau^*(x_1,x_2) \leq 0\, . 
    \label{triangle}
\end{split}
\end{equation}
Note that here $\Delta_\tau(x_1,x_2|x_0)= \Delta_\tau(x_2,x_1|x_0)$, due to the symmetry of diffusion the dynamics, but asymmetric dynamics such as advection on directed graphs can also be considered, see Section~\ref{section_celegans}.
Indeed, if $x_1$ and $x_2$ are near the opposite boundaries of the segment, they do not reach each other. Yet a point $x_0$ in the center (between them) will reach both, leading to a violation of the triangle inequality.
Hence if a source point $x_0$ is a participant in a high number of violations of the triangle inequality~\eqref{triangle}, then it is highly central. 

This leads to our definition of the \emph{multiscale centrality} (MSC) for a point $x_0$ parametrically dependent on the scale $\tau$. The MSC is the fraction of all pairs of points $(x_1,x_2)\in [0,1]^2$ that violate the triangle inequality \eqref{triangle} relative to $x_0$ over time horizon $\tau$
\begin{align}
\label{eq:definition_MSC}
    \mathcal{M}_\tau(x_0) = \Big \langle \mathbf{1}_{\Delta_\tau(x_1,x_2|x_0) \leq 0} \Big \rangle\, , 
\end{align}
where $\mathbf{1}_A$ is the indicator function for event $A$ and $\langle \cdot \rangle$ denotes the average over the cartesian product $[0,1]^2$.
Figure~\ref{fig:main_figure}(c) shows $\mathcal{M}_\tau(x_0)$ 
for $x_0\in[0,1]$ as a function of the horizon $\tau$.
As expected, the MSC is large at the midpoint of the segment and decreases toward the boundaries for scales $\tau$ that are large enough so that the dynamics can feel the boundaries and `locate' the centre of the segment with respect to its ends. 
In the absence of boundaries, the MSC is constant for the infinite line (i.e., $\mathcal{M}_\tau(x_0)=m(\tau), \enskip \forall x_0$) since there are no violations of the triangle inequality. Hence in that case, we find no center. 
The MSC thus allows us to establish the center of the domain, relative to boundaries of the underlying space, from the observation of the diffusion dynamics.

This notion of geometry based on diffusive dynamics is defined not only with respect to the domain boundaries, but also takes into account the effective reachability modulated by any inhomogeneities in the domain. Indeed, although we started by considering the simple case of a diffusion~\eqref{diffusion-segment} with a constant coefficient $D$, the same procedure applies for a non-homogeneous diffusion.  In that case, MSC incorporates information about the mass distribution at different scales, i.e., it is analogous to finding scale-dependent centers of mass, as we discuss more extensively in Section "Centrality in irregular and noisy meshes" below.

\subsection{Extending the definition to graphs}

The main ingredient in our definition of multiscale centrality~\eqref{eq:definition_MSC} is the use of the diffusion dynamics to infer the centre of a (possibly inhomogeneous) compact space relative to its boundaries.   
Although in spaces described by graphs  the topological notion of boundary is not easy to establish, it is straightforward to generalize our geometric definition of centrality by invoking a diffusive dynamics on the graph. 

Consider an undirected, connected graph with $N$ nodes, (weighted) adjacency matrix $A$, and degree matrix $K = \mathrm{diag}(A \bm{1})$. For simplicity, we concentrate on undirected graphs initially and present the case of directed graphs separately at the end of the paper. 
The definition of MSC then translates directly to graphs by choosing the diffusive process
\begin{align}
    \partial_t \mathbf p(t) = -L \mathbf p\, , 
    \label{diff-network}
\end{align}
where $L = K-A$ is the graph Laplacian, and the $N \times 1$ time-dependent node vector $\mathbf p(t)$ takes the place of the function $p(x,t)$ in~\eqref{diffusion-segment}. For an initial condition with a delta function at node $i$, the $j$-th coordinate of the solution of~\eqref{diff-network} is given by
\begin{align}
    p_j(t|i) = \left (e^{-tL}\right)_{ij}\, .
    \label{eq:network_solution}
\end{align}
Note that in all subsequent calculations, we rescale time by the spectral gap of the Laplacian (i.e., $t=t/ \lambda_2$) to have a comparable time scale across graphs.

Following closely the geometric derivation above, Figure~\ref{fig:main_figure}(d)-(f) illustrates the application to graphs through the Zachary Karate club~\cite{Zachary1977}, a social network that has been widely studied as it underwent an acrimonious split into two factions led by the `Officer' and `Mr Hi'.
A snapshot of the solution~\eqref{eq:network_solution} with an initial delta function on `Mr Hi' is shown in Figure~\ref{fig:main_figure}(d), where we also see that some of the node functions $p_j(t|i= \text{`Mr Hi'})$ exhibit a peak (e.g., the Bridge node) whereas other nodes (e.g., the 'Officer') do not overshoot~\cite{bacik2016flow}.

We can thus compute the peak time $t^*_{ij}$ at which a peak appears at node $j$ when an impulse is injected at node $i$ (with the convention $t^*_{ij}=\infty$ if there is no peak), as shown in Figure~\ref{fig:main_figure}(e). 
As expected, the two known clusters of the Karate club (associated with the split between `Mr Hi' and 'Officer') are separated by their reachability, yet there is a group of nodes that bridge across both clusters where typically misclassified nodes fall. 

We then follow the previous construction of \textit{reachability within a time horizon $\tau$} defined for the line segment in~\eqref{eq:peaktimes_tau}, and compute the scale dependent functions $t^*_{ij, \tau}$ for the graph.
There will be pairs of nodes $(j,k)$ for which the triangle inequality centered at node $i$ is violated, so that 
\[ \Delta_{ij,\tau} :=  t^*_{ij, \tau} + t^*_{ik, \tau} - t^*_{jk, \tau} \leq 0, \] 
where $t^*_{ij, \tau}$ is defined as in~\eqref{eq:peaktimes_tau}.
Adapting the continuous definition~\eqref{eq:definition_MSC} to the discrete setting, we can thus compute $\mathcal{M}_\tau(i)$, the MSC of node $i$ at scale $\tau$, as the proportion of violations of the triangle inequality in which $i$ is involved up to time $\tau$, and 
summarize this information through the normalized multiscale centrality node vector $\bm{\mathcal{M}}_\tau$ with elements
\begin{align}
    (\bm{\mathcal{M}}_\tau)_i = \mathcal{M}_\tau(i)/\max_j\, \mathcal{M}_\tau(j) .
    \label{eq:norm_MSC}    
\end{align}
Fig.~\ref{fig:main_figure}(f) shows the multiscale centrality  $\bm{\mathcal{M}}_\tau$ at three different scales $\tau$ for all nodes of the network. The measure shows that both Mr. Hi and the Officer are central nodes at small scales (i.e., they are central to their local environments), whereas the nodes bridging across both clusters become central at larger scales (i.e., they are central relative to the full network).

\section{Applications}
\subsection{Centrality in small social networks: from high degree to closeness}

\begin{figure*}[htpb!]
    \centering
    \includegraphics[width=0.9\textwidth]{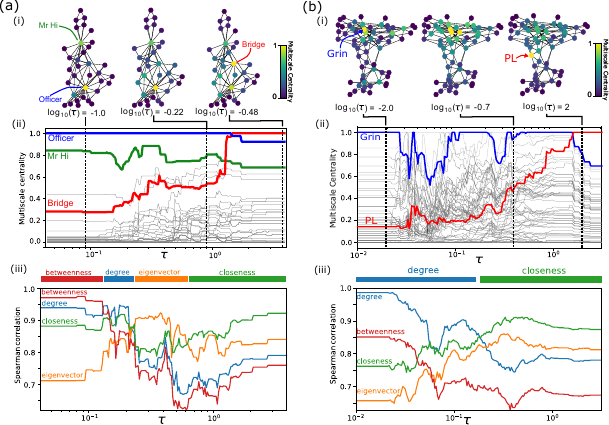}
    \caption{\textbf{ Multiscale centrality of (a) the Karate club network and (b) the Dolphin network.}~\cite{Lusseau2007}
    (i) The multiscale centrality \Mv~\eqref{eq:norm_MSC} for short, middling and long scales.
    (ii) The multiscale centrality \Mv for all nodes as a function of the scale $\tau$.
    (iii) Spearman correlation between \Mv and four common centrality measures as a function of the scale $\tau$. \Mv is most correlated with betweeness centrality for Karate club network and degree centrality for Dolphin network respectively at small scales, and both are most correlated with closeness at large scales. 
    }
    \label{fig:examples}
\end{figure*}

In addition to the results on the Karate club networks of Figure~\ref{fig:main_figure}
(d)-(f), we show in Figure~\ref{fig:examples}(a) the ranking of nodes across all $\tau$, highlighting the most central nodes at three different scales. 
The two leaders are most central across a wide range of scales, however, at long scales we find that a third node (that bridges the two groups) becomes the most central. 
The Spearman correlation shows that the ranking induced by multiscale centrality is most strongly correlated with betweenness at short scales due to the particularities of the structure and size of the Karate club network. 
We observe a high correlation with eigenvector centrality at middle scales and closeness at large scales. We note that across all scales Mr Hi and Officer are highly central (corroborated by the four standard centrality measures), suggesting that these two nodes are central independent of scale.  
From this simple example, we see that nodes with high degrees (`hubs') are central at small scales, whereas `inter-hub nodes' are central at larger scales relative to the global structure of the graph. This observation reflects the different heuristics that have been used in the literature to define centrality, from the local to the global. 

To further examine the correspondence of our measure with other centralities, we have analysed in Figure~\ref{fig:examples}(b) a second classic example, the interaction network of bottlenose dolphins, constructed between $1994$ and $2001$ in New Zealand~\cite{Lusseau2007}. 
Figure~\ref{fig:examples}(b)i displays $\bm{\mathcal{M}}_\tau$, the multiscale centrality for three values of the scale $\tau$.
At small scales, the dolphin Grin (with the highest degree) has the highest centrality whereas at large scales another dolphin, called PL,  becomes the most central, highlighting its role as a connector.  
Interestingly, both dolphins exhibit rare communication traits~\cite{Lusseau2007}.
A minority of the dolphins within the dataset displayed physical forms of communication to other dolphins; 7/63 dolphins display `lobtailing' and 5/63 display `side flopping', which are both forms of physical to auditory communication. The dolphin Grin is located at the centre of a dolphin cluster and is one of the few dolphins that displays lobtailing communication. Contrary to Grin, PL, the central node at large timecales, displayed `side flopping' communication which suggests that, despite its low degree, PL communicates across the two main clusters of  dolphins. 
Figure~\ref{fig:examples}(b)iii shows the Spearman correlation of the ranking based on $\mathcal{M}_\tau(i)$ compared to four classical notions of centrality. 
The multiscale centrality at low $\tau$ is highly correlated with the degree, whilst at large $\tau$ we observe a strong correlation with closeness. 
This result recovers our previous results on Karate club network: the degree is a local measure whereas closeness takes into account distances to all nodes in the network.
Betweenness shows similar trends to degree centrality and eigenvector centrality shows similar behaviour to closeness, albeit they do not correlate as strongly with multiscale centrality and tend to correlate at various intermediate scales depending on the graph.

\subsection{Centrality from heterogeneity: random and irregular meshes}
\label{sec:Delaunay}

The proposed notion of centrality is related not only to the geometric center but, more generally, to the \emph{center of mass}, i.e., it incorporates information from both boundaries and local density. We illustrate this fact through the analysis of (noisy) Delaunay meshes on the plane. 

\begin{figure}[htpb]
    \centering
    \includegraphics[width=0.45\textwidth]{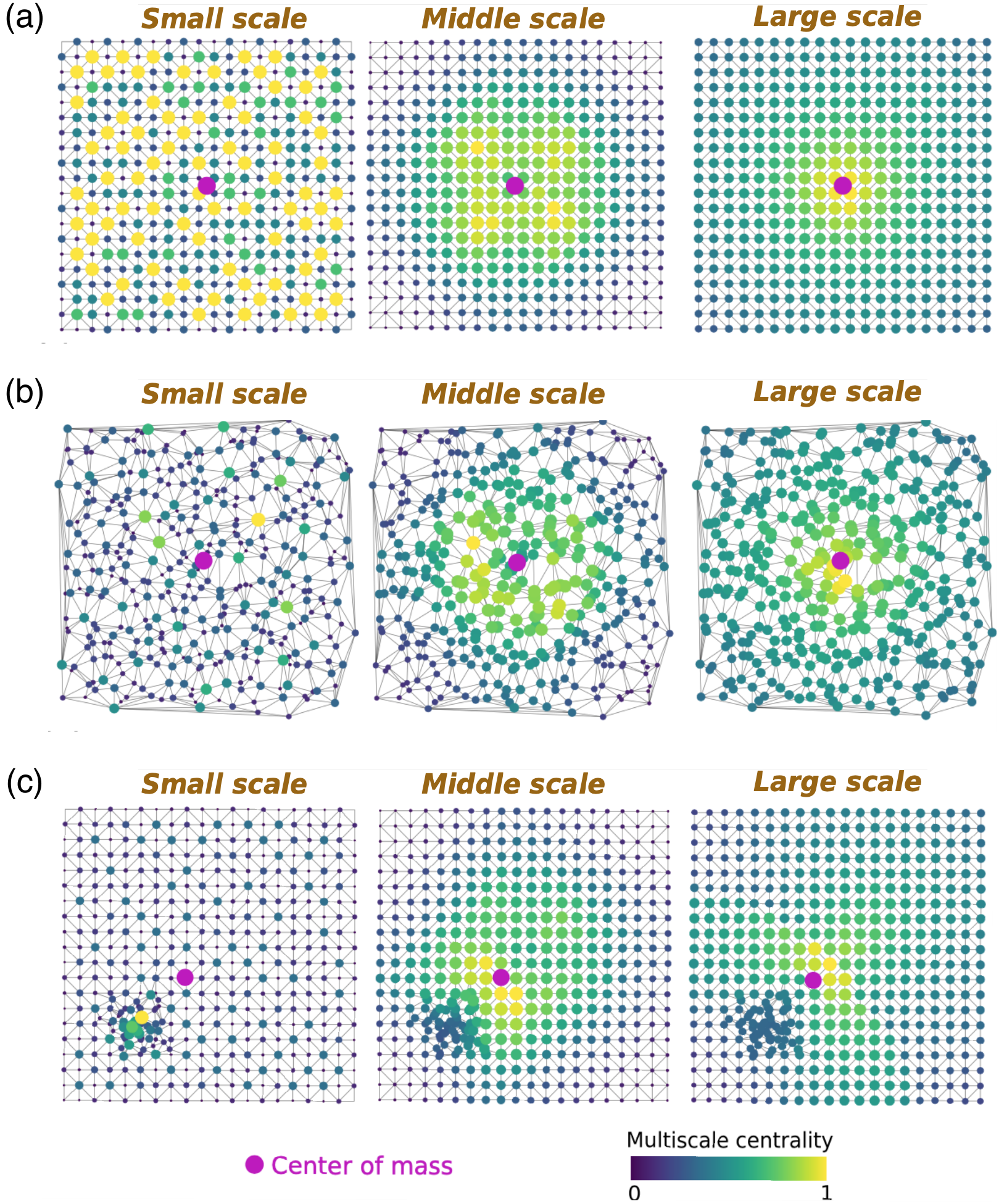}
    \caption{
    \textbf{Multiscale centrality of weighted graphs obtained from Delaunay mesh triangulations on the plane.} 
    (a) Uniform grid: At small scales, \Mv is highly correlated with degree, whereas at large scale \Mv coincides with the center (geometric centre and centre of mass), thus correlated with closeness. 
    (b) Noisy uniform grid obtained by adding a Gaussian random deviation to the positions on the grid with variance $0.03$. At small scales, the centrality again reflects the degree of the nodes and at large scales, the geometric center and center of mass of the square is still well captured by \Mv. 
    (c) Inhomogeneous mass distribution obtained by adding to the grid a dense patch of nodes Gaussianly distributed around the point $(0.2,0.2)$. At small scales, \Mv correlates with the degree, with high localization at the dense added patch of nodes.
    At large scales, the centrality \Mv is concentrated on nodes between the center of mass and the geometric center of the square due to the combined effect of the mass distribution and the effect of the domain boundaries.}
    \label{fig:grid_delaunay}
\end{figure}

Figure~\ref{fig:grid_delaunay}(a) shows the analysis of the Delaunay mesh of a regular (non-random) $20 \times 20$ grid discretization of the flat unit square $[0,1]^2$, 
where each edge $ij$ has weight $w_{ij} = 1/l_{ij}=1/ \|x_i-x_j\|$, 
where $x_i$ and $x_j$ are the positions of nodes $i$ and $j$, respectively.
%
We compute the multiscale centrality \Mv of this weighted graph across scales $\tau$. As expected, at small scales \Mv correlates strongly with the weighted degree of the node, whereas at the middle scales, the centrality mirrors the presence of the square boundaries. At large scales  \Mv is peaked at the center of the square. This behaviour is similar to the one-dimensional case in Fig.~\ref{fig:main_figure}(a)-(c).  
We also computed the center of mass of this regular network, where the mass of each node is its weighted degree. Since the mass distribution is close to uniform, the center of mass corresponds to the geometric center, and is well recovered by the centrality at large scales. 

To test the robustness of these results under randomness,  we analyzed the weighted network built from the Delaunay mesh of a noisy grid, where the grid positions $x_i$ have a Gaussian standard deviation of $0.03$ (Fig.~\ref{fig:grid_delaunay}(b)). 
Again, we find that \Mv correlates t small scales with the node degree, picking the local inhomogeneities created by the random fluctuations. At large scales, we find that \Mv recovers the geometric center, which is also the center of mass in this case, even for such a complex and non-uniform discretisation of the unit square.

To study the effect of larger deviations from the uniform grid, we studied in Fig.~\ref{fig:grid_delaunay}(c) another inhomogeneous Delaunay mesh obtained by adding $50$ additional nodes on the plane, drawn from a normal distribution centered at $[0.2,0.2]$ with variance $0.05$.  This amounts to adding a localized mass in the bottom left quarter of the square, thus introducing a non-uniform mass distribution. 
The computation of \Mv for the corresponding weighted network shows that, at small scale, the centrality captures the higher degree (i.e., the mass) of the added cluster of dense nodes, whereas at large scales, the most central nodes are situated close to the centre of mass, which lies between the added cluster and the geometric center of the square. 
Note that the peak of the \Mv does not exactly match the center of mass due to the effect of the boundaries of the square domain. Hence the multiscale centrality peaks at a point between the geometric centre and the centre of mass.

\subsection{Application to geographically embedded networks: power grid and road networks} 
\begin{figure*}[htpb]
    \centering
    \includegraphics[width=1.0\textwidth]{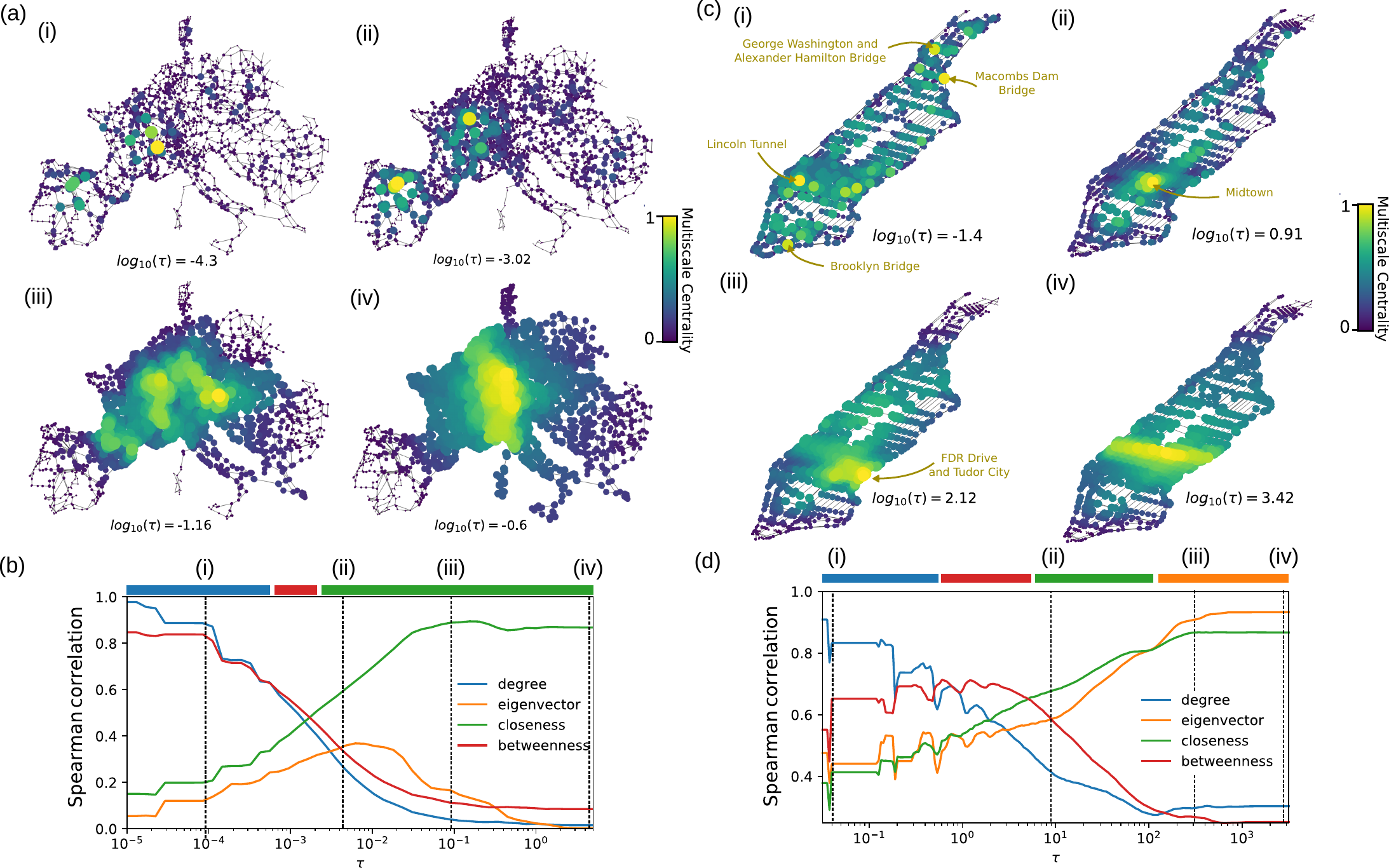}
    \caption{\textbf{Multiscale centrality of the European power grid network ~\cite{schaub2012markov} and Manhattan road network~\cite{boeing2017osmnx}.}
    (a) The European power grid exhibits a multicentric structure, which is revealed at different scales. The centrality \Mv is shown at four increasing scales (i)-(iv). At small scales, local centers are found in Spain and France, then progressive larger nuclei emerge in south west France, the Paris region, the Pyrenees coalescing onto 
    a strip stretching from Venice to Amsterdam, with a large centrality region in Germany and Switzerland. 
    (b) The Spearman correlation of \Mv with four classic centrality measures for the European power grid. (c) The Manhattan road network at four increasing scales (i)--(iv). At small scales, the major bridges and tunnels into Manhattan are most central, followed by a coagulation of centrality in midtown which then progressively shifts towards the east coast before a final central region appears stretching between midtown west and midtown east. (d) The Spearman correlation of \Mv with four classic centrality measures for the Manhattan road network.
    }
    \label{fig:powergrid_main}
\end{figure*}

In larger, complex networks linked to geometric settings, $\bm{\mathcal{M}}_\tau$ can reveal the multiscale nature of the network centrality. An example is given in Fig.~\ref{fig:powergrid_main}(a)-(b), where we study the (unweighted) network of the European power grid (data from the Union for the  Coordination of Transmission Energy, see~\cite{schaub2012markov} for more details on this dataset). 
This network contains $2783$ nodes and $3762$ unweighted edges, with community structure present at several scales, see~\cite{schaub2012markov}.

Figure~\ref{fig:powergrid_main}(a) displays four instances of the multiscale centrality \Mv of the power grid computed at different scales $\tau$.
At small scales (i) $\log_{10}(\tau) = -4.3$, \Mv correlates strongly with degree centrality and the most central nodes are highly local: the most central node is in the south west of France, followed by various cities located predominantly in France and Spain, suggesting that these countries contain nodes with a higher local centrality (relative to the rest of Europe), maybe due to some systematic difference in the design of the power grid in those countries. At slightly longer timescales (ii) $\log_{10}(\tau) = -3.02$, we find the most central nodes located predominantly in the north of Spain and near Paris in France, areas with a mixture of high degree and bridging across domains.
At yet longer timescales (iii) $\log_{10}(\tau) = -1.16$, the nodes in Spain become less central, reflecting their more peripheral location relative to the global network, and instead we observe three main regions of high \Mv: a region around the Pyrenees, on the border between Spain and France; a region along the border between France and Germany; and a region on the western part of Eastern Europe. At the largest timescales (iv) $\log_{10}(\tau) = 0.6$, we find a single predominant region of highly central nodes stretching from the north-east of France to the north-west of Italy, where the east and west centers present at shorter scales collapse. 
Figure~\ref{fig:powergrid_main}(b) shows the Spearman correlation of  \Mv with four classic centrality measures. As observed above, \Mv correlates well with degree and betweenness at small scales and with closeness at large scales. 

To further illustrate the application of multiscale centrality to such engineering networks, we have constructed the Manhattan road network using the \textsc{osmnx} python package~\cite{boeing2017osmnx} without residential roads to reduce network size. 
Similarly to the analysis of the power grid, Figure~\ref{fig:powergrid_main}(c) displays \Mv computed at four time scales and mapped onto the road network.
At small scales (i) $\log_{10}(\tau) = -1.4$, where \Mv correlates mostly strongly with degree, we find that nodes associated with major entrances onto the island are most central, including the Lincoln tunnel and the joint connection of George Washington and Alexander Hamilton bridge. This is expected since the major bridges and tunnels into Manhattan usually have multiple road connections (high degree) to facilitate the flow of traffic. At middling timescales (ii) $\log_{10}(\tau) = 0.91$ we identify midtown as the most central, an area that attracts tourists and major companies. As the scale increases (iii) $\log_{10}(\tau) = -2.12$, we find that the centre of the network shifts towards FDR Drive on the east coast of the island. FDR drive is a major highway that connects distal regions of the island. 
Finally, at long timescales (iv) $\log_{10}(\tau) = 3.42$ we identify an East to West band from $54$th to $59$th street between $4$th Avenue and $8$th Avenue as most central.
As above, Figure~\ref{fig:powergrid_main}(d) shows the Spearman ranking of $\bm{\mathcal{M}}_\tau$ with a good alignment of \Mv with degree at small scales and with closeness and eigenvector centrality at large scales. 

These analyses thus indicate a multi-centric structure in both the European power grid and Manhattan road network, highly dependent on the scale of interest, which reflects design principles and suggests that the proper scale should be considered when establishing the importance of nodes within such complex networks.

\subsection{Application to directed graphs:  the neuronal network of C.\ elegans}\label{section_celegans}

Finally, we consider the application of multiscale centrality to directed networks. The definition of \Mv in terms of a diffusion process lends itself naturally to incorporate the effects of directionality in networks. We illustrate this point through the analysis of  the directed and weighted graph of the neuronal connectome  of the nematode \textit{C.\ elegans}~\cite{bacik2016flow}.

For directed graphs $A \neq A^T$, and we need the following modification of the definition of the multiscale centrality. Consider the directed combinatorial Laplacian with teleportation~\cite{lambiotte2014random,Schaub2018d}:
\begin{align*}
    L_{dir} = \Phi - \frac12 \left(\Phi P + P^T \Phi \right )\, , 
\end{align*}
where $\Phi = \text{diag}(\bm{v}_1)$ contains on the diagonal the Perron (leading) vector $\bm{v}_1$ of the transition matrix with teleportation 
\begin{align*}
    P = \alpha D^{-1} A + \Big( (1- \alpha) + \alpha\, \mathrm{diag}(\bm{a})\Big )\frac{\bm{1} \bm{1}^T}{n}\, . 
\end{align*}
Here $\alpha = 0.85$ is the (Google) teleportation parameter; $\bm{a}$ is an indicator vector function for nodes with vanishing out-degrees; and $\bm{1}$ is the vector of ones. 

For a directed graph, $t^*_{ij}\neq t^*_{ji}$ in general. We thus consider the following extension of the triangle inequality 
\begin{align}
    t^*_{ij} + t^*_{ik} \geq \frac{t^*_{jk}+ t^*_{kj}}{2}\, ,
\end{align}
and again count the fraction of violations of this inequality for any node $i$ as the basis to define our centrality measure.
This directed version of the multiscale centrality is therefore sensitive to the directionality of the graph.

\begin{figure}[h!]
    \centering
    \includegraphics[width=1.0\columnwidth]{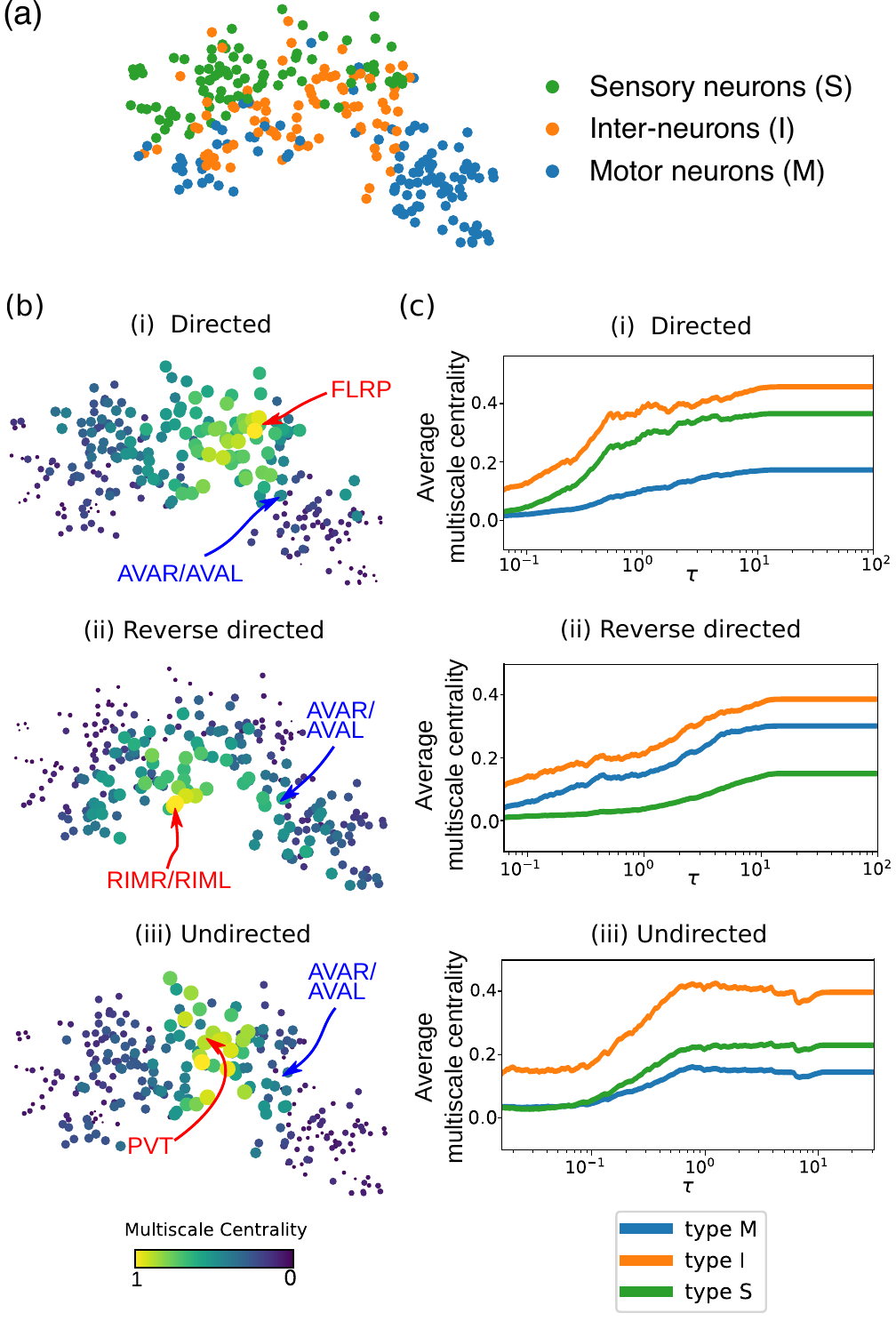}
    \caption{Multiscale centrality of the irected and and weighted neuronal network of \textit{C.\ Elegans}. 
    (a) The neuronal network of \textit{C.\ Elegans} has three types of neurons: sensory (S), inter (I), and motor (M) neurons.
    We compute \Mv for: (i) the original directed network; (ii) the reverse directed network; and (iii) the undirected network where the directionality of edges is ignored.  
    (b) We show \Mv for each of those networks at large values of the scale $\tau$. The most central nodes at large scales for each of the networks are: (i) FLPR (a sensory neuron); (ii) RIMR/RIML (two motor neurons); (iii) PVT (interneuron) and AQR (sensory neuron). The most central nodes at small scales $\tau$ for all three graphs (in blue) are the AVAR/L neurons.
    (c) The multiscale centrality averaged over the neurons of each type (S,I, M) is computed for the three networks. Interneurons are most central for all three networks at all scales. Sensory neurons only become central at large scales for the directed network, whereas motor neurons only become central at large scales for the reverse directed network. In the undirected network, neither motor nor sensory neurons are central at any scale. 
    }
    \label{fig:celegans}
\end{figure}

As an example, we have analyzed the directed neuronal network of \textit{C. elegans}. 
This network has three types of neurons: sensory (S), interneurons (I), and motor (M)
(Fig.~\ref{fig:celegans}(a)). Broadly speaking, S neurons process environmental stimuli, which are channelled downstream through I neurons towards the M neurons in charge of effecting motion as a reaction to external inputs. Hence we expect an overall directional flow of information from S to I to M neurons.  

To explore the effect of directionality, we have computed and compared the multiscale centrality \Mv of the directed \textit{C. elegans} network, its reverse, and an  undirected version where directionality is ignored. The results are presented in Fig.~\ref{fig:celegans}(b)-(c).
We find that in all three networks (directed, reverse directed and undirected) the AVAR and AVAL neurons are consistently the most central nodes at short timescales. This is a consequence of AVAR and AVAL having both the highest in and out degrees of any nodes in the network. 
At larger scales, however, there are differences between the three networks: in the directed network, FLPR (a sensory neuron) is the most central, whereas in the reverse directed network, RIMR/RIML (two motor neurons) become the most central. In the undirected network, we find PVT (inter neuron) and AQR (sensory neuron) as the most central nodes at large scales.
In general, we find that the most central nodes in the directed network are interneurons followed by sensory neurons (and low \Mv for motor neurons), whereas 
the most central nodes in the reverse directed network are interneurons followed by motor neurons (with low \Mv for sensory neurons). In the undirected connectome, interneurons are the most central, whereas both sensory and motor neurons have low centrality. This is shown in Fig.~\ref{fig:celegans}(c), where we compute the average centrality for each of the three classes of neurons. These results show the importance of including directionality and scale in the analysis of directed networks.

\section{Conclusion}

We have introduced a scale-dependent graph centrality, 
which is based on a notion of reachability of nodes from a localized diffusive source in terms of overshooting events. 
The key concept is to interpret the timing of the overshooting events as a proxy for a distance between nodes, such that the underlying geometry of the network is captured by the diffusive process. Within this framework, central nodes are those that are involved in breaking the metricity of the diffusion.
The proposed measure recovers the concept of a center of mass relative to the intrinsic boundaries and density inhomogeneities of the graph probed by the diffusion. Because the diffusion has an inherent dynamical scale, our measure captures different notions of centrality, from the local (degree) to the global (closeness). We have illustrated the multiscale centrality measure through social networks, geometric networks and geographically-embedded networks.  

A Python code to compute the multiscale centrality~\eqref{eq:norm_MSC} is available at \url{https://github.com/barahona-research-group/MultiscaleCentrality}. In its current form, the method is applicable to relatively large graphs (thousands of nodes) but the evaluation of the matrix exponential~\cite{al2011computing} and the triangle inequalities can become expensive for larger graphs. 
To extend this measure to larger graphs, the matrix exponential could be approximated or other centrality measures could be extracted by, e.g., treating the matrix of peak times $t^*_{ij}$ as a distance matrix and computing a different version of the closeness centrality.

The dynamical foundation of the \Mv measure means that directed graphs can also be considered seamlessly within this approach. We provide an example of directed graphs, where we study the multiscale centrality of the neuronal network of \textit{C.\ Elegans}, which captures differences between forward, backward and symmetrized graphs related to biological information flow from sensory to motor neurons (Fig.~\ref{fig:celegans}). 

Finally, we remark that although we have used here a diffusive process for conceptual clarity, similar centrality notions could be defined to incorporate more complex dynamics, such as epidemic spreading, Kuramoto oscillators, or non-Markovian dynamics, and may provide new tools to study such dynamics from a diffusion-based perspective~\cite{zhang2011node}.

\subsection{Acknowledgments} 
We thank Paul Expert, Tarik Altuncu, Asher Mullokandov, Isabel Ashman, Karol Bacik, Michael Schaub and Sophia Yaliraki for valuable discussions. 
We acknowledge funding through Engineering and Physical Sciences Research Council award EP/N014529/1 supporting the EPSRC Centre for Mathematics of Precision Healthcare at Imperial.

\bibliography{refs}    

\end{document}